# Optical Damage Threshold and THz Generation Efficiency of (Fe,CoFeB)/(Ta,Pt) Spintronic Emitters

Sandeep Kumar[1], Anand Nivedan[1], Arvind Singh[1], Yogesh Kumar[2], Purnima Malhotra[2], Marc Tondusson[3], Eric Freysz[3], and Sunil Kumar[1,*]

[1]*Femtosecond Spectroscopy and Nonlinear Photonics Laboratory, Department of Physics, Indian Institute of Technology Delhi, New Delhi 110016, India*
[2]*Laser Science and Technology Center, Metcalfe House, Civil Lines, New Delhi 110054, India*
[3]*Univ. Bordeaux, CNRS, LOMA, UMR 5798, 33405 Talence, France*
*Corresponding Author: kumarsunil@physics.iitd.ac.in

THz pulses are generated from femtosecond pulse-excited ferromagnetic/nonmagnetic spintronic heterostructures via inverse spin Hall effect. The contribution from ultrafast demagnetization/remagnetization is extremely weak, in the comparison. The highest possible THz signal strength from spintronic THz emitters is limited by the optical damage threshold of the corresponding heterostructures. The THz generation efficiency does not saturate with the excitation fluence even up till the damage threshold. Bilayer (Fe, CoFeB)/(Pt, Ta) based FM/NM spintronic heterostructures have been studied for an optimized performance for THz generation when pumped by sub-50 fs amplified laser pulses at 800 nm. Among them, CoFeB/Pt is the best combination for an efficient THz source. The optimized FM/NM spintronic heterostructure on a quartz substrate, having α-phase Ta as the nonmagnetic layer, show the highest damage threshold as compared to those with Pt, irrespective of their generation efficiency. The damage threshold of the Fe/Ta heterostructure on quartz substrate is ~85 GW/cm$^2$.

## 1. Introduction

When ferromagnetic/nonmagnetic (FM/NM) metallic thin film heterostructures are irradiated with femtosecond laser pulses, emission of THz pulses takes place.[1-7] Such emitters are unique in the sense that they combine the excitements from three most active research fields, currently, the ultrafast lasers, spintronics, and THz radiation. The most commonly known process responsible for the THz pulse generation from FM/NM type spintronic heterostructures is the inverse spin Hall effect (ISHE).[1-7] Beaurepaire *et al.*, for the first time in 2004, observed emission of THz radiation from femtosecond laser excited ferromagnetic (Ni) thin film, where, it was suggested that femtosecond laser-induced ultrafast demagnetization process was responsible for the THz pulse emission.[8] From the subsequent studies[9] it was realized that only a weak THz radiation is possible from such a process. The role of the ultrafast demagnetization process for THz generation from single-layer FM films was further investigated with respect to the thickness of the film to find an optimal efficiency for a specific thickness.[10] Zhang and coworkers attributed anomalous Hall effect for the generation of THz radiation from thin single ferromagnetic, $(Fe_xMn_{1-x})_yPt_{1-y}$ layers.[11] Similarly, a few other studies in the literature have indicated the generation of a very weak THz radiation from femtosecond laser excited NM metal based structures.[12] The issue of low efficiency was quickly resolved as soon as FM/NM-based metallic heterostructures were experimented upon using the femtosecond laser excitation,[7] where, the ISHE was considered to be the underlying mechanism.[13] Not only a high generation efficiency but also broad bandwidth of the generated THz radiation became realizable.[6] Here, the spin Hall angle in the NM heavy metal layer, the spin polarization in the FM layer, the individual film quality, and thicknesses in the heterostructure, are the major contributing material parameters. In principle, the bandwidth of the THz radiation generated from these sources is limited by the duration of the excitation pulse only.[6]

THz generation from the above mentioned spintronic heterostructures shows a nonsaturating behavior, i.e., nearly linearly increasing THz power with respect to the excitation power.[2, 4, 6, 14, 15] By using proper thicknesses and combinations of the FM/NM layers in their heterostructures, and the optical conditions in the experiments, THz electric field strengths of a few hundreds of kV/cm along with broad bandwidth are achievable.[4] Conventionally, high power THz radiation is generated from amplified laser pulse-excited nonlinear crystals such as lithium niobate,[16] organic materials,[17] and the dual color air plasma.[18-21] For realization of high-power THz radiation from air plasma, it requires very high-energy pump-pulses. Nonlinear optical materials like lithium niobate crystal and organic crystals, which can sustain hard pumping and provide intense THz pulses, have limitations in terms of strict phase matching condition, absorption at THz frequencies, etc.[22] In fact, THz radiation with its electric field strength as high as a few MV/cm, is also possible these days.[23] In principle, the efficiency of any solid material based THz source is limited ultimately by the associated optical damage threshold.

The spintronic THz sources, very often, have been tested with low pulse energy femtosecond lasers for achieving broad bandwidth.[6, 14] In a very few cases, they have also been tested with the amplified laser pulses,[4] for achieving high power THz radiation. Beyond a certain value of the excitation power, spin accumulation in the NM layer contributes to a weak saturating behavior.[5, 7] Furthermore, like the nonlinear crystals, spintronic THz sources can be used only before their optical damage threshold. As compared to others, spintronic based THz emitters have a large scope in terms of the choice of the materials that can be used, better tunability in polarization, bandwidth, and power. Optimization through excitation wavelength, sample geometry, applied magnetic field, etc., can be achieved easily.[2, 4, 6, 14, 15, 24-28] In the current study, we have experimentally determined the optical damage threshold of many popular spintronic THz emitters on different substrates by using amplified laser pulses at 800 nm, which has not been reported explicitly in the literature, hitherto. Fe and CoFeB have been used for the FM layer and Pt and Ta (in α-phase) have been used as the NM layer in the bilayer spintronic heterostructure. These heterostructures have been tested for optimal THz power generation by using various combinations of the FM and NM layer thicknesses. Ultrafast demagnetization and inverse spin Hall effect have been explored for the THz generation mechanism. For Fe/Ta, the optical damage threshold is the highest, ~85 GW/cm$^2$, which is quite comparable with that of the popularly used lithium niobate crystal. The THz generation efficiency of the thickness optimized Fe/Pt structure is found to be even better than the dual color air plasma-based THz source for excitation pump energy levels below ~500 μJ. While, the CoFeB/Pt is more efficient source than Fe/Pt, being consistent with the literature, we find that THz power generation efficiency of Fe/Ta is much higher than that of the CoFeB/Ta.

2. Results

Figure 1 summarizes the outcomes on the THz generation efficiency from various thickness combinations in the bilayer FM/NM heterostructures on quartz substrates that are fabricated by magnetron sputtering in ultrahigh vacuum (see the Experimental Section). FM layer is of either Fe or CoFeB and the NM layer is of either Pt or Ta. The Ta layer is grown in its α-phase (see Supplementary Section S1). These results provide the optimized thicknesses and their combinations for the total thickness of the bilayer structure such that maximum THz power is produced from them under the same conditions with the optical excitation, room temperature and humidity, optical alignment, and the detection by electro-optic sampling. A schematic of the THz pulse generation, following excitation by a femtosecond laser pulse at near infrared (NIR), is shown in Figure 1(a). A representative THz time-domain signal, E(t) and its Fourier spectrum, E(ω) are presented in Figure 1(b) and 1(c), respectively, generated from Fe(3)/Pt(2) bilayer heterostructure at a reasonable value of the excitation pulse energy of ~0.3 mJ or fluence of ~1.2 mJ/cm$^2$. All the THz generation and detection measurements for this paper have been done under room temperature and humidity (~50%) conditions that are realistic for all practical purposes, mainly, the applications in stand-off detection and screening. Hence, the strong far IR absorption lines due to the characteristic water absorption bands[29] in the spectrum can be seen from Figure 1(c). The limited THz bandwidth (~0.2-5 THz) is mainly due to ZnTe crystal used in our electro-optic sampling setup. Also, if dry air is purged to remove the humidity down to a few presentage, we noticed about 10 times increase in the THz time-domain signal, however, this was not done for all the measurements and hence not discussed in the present paper.

The femtosecond pump excitation (z-direction) of the FM layer, which is kept magnetized by B$_{ext}$ along -y-direction, produces a spin current. The spin current, J$_s$ in the z-direction gets injected into the NM layer of the heterostructure. Due to strong spin-orbit coupling in the NM layer, by virtue of the ISHE,[13] a charge current, J$_c$ is produced in the x-direction as shown in Figure 1(a). Note that the magnetization direction is defined by the external magnetic field, B$_{ext}$, which is kept just above the saturation magnetization field of the FM layer. Briefly, the THz pulse generation process through ISHE in a FM/NM bilayer spintronic heterostructure is described here. NIR femtosecond excitation pulse stimulates the electrons in the FM metal layer from states below the Fermi level to above it, creating a non-equilibrium hot electron distribution.[30] The FM metal is such that there is a difference in its spin-up and spin-down electron densities and their transport properties in the respective bands.[31] This difference helps in producing a spin-polarized current (J$_s$) to be injected into the NM layer by means of super diffusive process if appropriate thickness of the FM layer is used.[30, 32] Since, the NM layer of the spintronics heterostructure is a heavy metal with high intrinsic spin-orbit coupling, opposite polarity spins are deflected in opposite directions by an amount proportional to the spin-Hall angle (γ) and hence produce a charge current density,[33] given by the following relation.[2, 7]

$$\vec{J_c} = \gamma \cdot (\vec{J_s} \times \hat{m}) \qquad (1)$$

Here, $\hat{m} = \vec{M}/|M|$ is the magnetization direction. Therefore, the in-plane transient charge current J$_c$ in the NM layer generates a pulse of THz radiation which has been measured by electro-optic sampling in our experiments.

Figures 1(d) to 1(i) show the results obtained from Fe/Pt structures on quartz for different thickness of the two individual layers in nanometers (nm) as mentioned inside small parentheses. The raw THz signals obtained for fixed Fe thickness and varying Pt thickness are shown in Figure 1(d) and 1(e). Similarly, raw results for fixed Pt layer thickness and varying Fe layer thickness, are presented in Figure 1(f) and 1(g). The consecutive time-domain traces in Figure 1(d-g) have been shifted vertically for clarity. As can be seen from Figure 1(h) and 1(i) that the root mean squared (RMS) amplitude of the THz field is dependent on the thickness of the individual layers, there is an optimum value of the thickness where maximum THz power is generated. It is clear from the thickness-thickness color intensity image plot in Figure 1(i) that the optimum thickness of the Fe layer is ~2 nm, that of the Pt layer is ~3 nm and for other values of thicknesses of the two layers, the maximum THz generation is found to be for thickness ~5 nm of the combined structure. These results are consistent with the literature.[14] Therefore, for referral purposes later on in the manuscript, the optimized Fe/Pt heterostructure on quartz contains 2 nm thick Fe layer and 3 nm thick Pt layer. The optimum thickness of the Fe layer is very close to its critical thickness (~1.5 nm) for the change in the direction of magnetic easy axis in the layer (please see the Supplementary Section S1).

Various parameters such as spin current generation in the FM layer, spin to charge current conversion and spin accumulation in the NM layer, optical and THz absorption in the total thickness, etc., contribute to the overall THz generation efficiency of the bilayer FM/NM heterostructures. From Figure 1(h), it can be seen that for a fixed thickness of the Fe layer, there is an optimum thickness of the Pt layer at which maximum THz generation can be obtained. Efficient conversion of the spin-polarized current into the transient charge current in the NM layer is necessary for maximum THz generation efficiency[34]. The decrease in the THz signal at lower thicknesses is due to the limited spin decay length in the NM material. The optimum thickness of the Pt layer for generation of the highest THz signal from the heterostructure is about 3 nm, which is just above the spin decay length of ~1 nm in Pt.[35] The dramatic change in the THz RMS amplitude with the thickness of the Fe layer changing from 2 nm to 1 nm in Figure 1(h), is attributed to the change in magnetization easy axis going from being in-plane to out-of-plain, below a certain thickness (see the supplementary Section S1). From Figure 1(h) and 1(i), it is clear that the optimum thickness of the Fe layer is about 2 nm. At smaller thicknesses of the Pt layer, the THz power continues to grow with the increasing thickness up till its optimum value of ~3nm. At the optimum thickness of the Fe layer, the decrease in the THz generation efficiency with the increasing Pt layer thickness (beyond 3 nm) can be understood in terms of a compromise between the spin to charge current conversion for the fixed pump fluence and the overall THz attenuation in the bilayer heterostructure[6]. The former is due to the spread of absorbed pump power over larger thickness in thicker films. The overall thickness beyond which the THz generation efficiency starts to decrease is ~5nm.

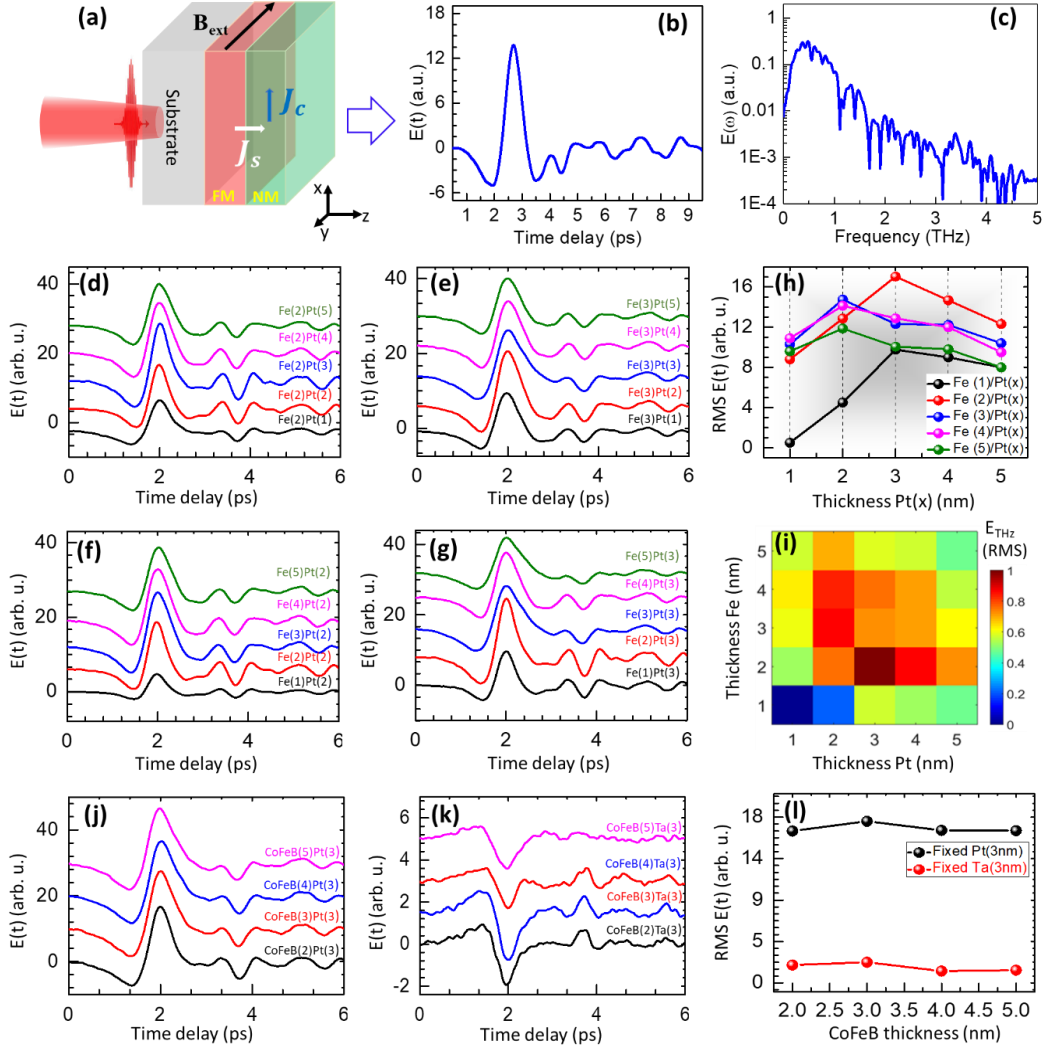

**Figure 1.** THz pulse generation from femtosecond NIR (800 nm) pulse irradiated FM/NM bilayer-type spintronic emitters deposited on quartz substrates at the excitation pulse energy (fluence) of ~0.35 mJ (1.2 mJ/cm$^2$). (a) Schematic of the optical pulsed excitation and THz pulse generation. $B_{ext}$ is the external magnetic field, $J_s$ is spin current and $J_c$ is charge current density. (b) Typical time-domain THz signal, E(t) detected by electro-optic sampling, and (c) the Fourier transformed spectrum, E(ω). (d-i) THz signals from Fe/Pt heterostructures for various combinations of thicknesses (nm) of the Fe and Pt layers. (j-l) THz signals from CoFeB/Pt and CoFeB/Ta bilayers for various thickness combinations of the two layers.

The raw THz results for fixed thicknesses of Pt or Ta layer at 3 nm while varying thickness of FM CoFeB layer from 2 to 5 nm in the two bilayers are presented in Figure 1(j) and 1(k), respectively. The RMS values of the THz signal for each sample are presented in Figure 1(l). It can be seen that the optimum value of the CoFeB layer thickness is close to 3 nm beyond which there is not much improvement in the THz generation efficiency and also the THz attenuation in the CoFeB does not increase much with its thickness upto 5 nm. In these cases, we have already used the optimum thickness of the Pt layer of ~3 nm. Our results (Figure 2) indicate that the optimum thickness of the Ta layer in the heterostructures is more than 5 nm, while almost no THz signal can be measured from the bilayers containing Ta layer thickness below 2 nm. In addition to the opposite polarity, we also find much lower THz generation efficiency of the CoFeB/Ta than the CoFeB/Pt bilayers (Figure 1(j,k)). Both of these effects can be attributed to the opposite sign and much smaller spin-Hall angle in the Ta as compared to Pt.[35, 36]

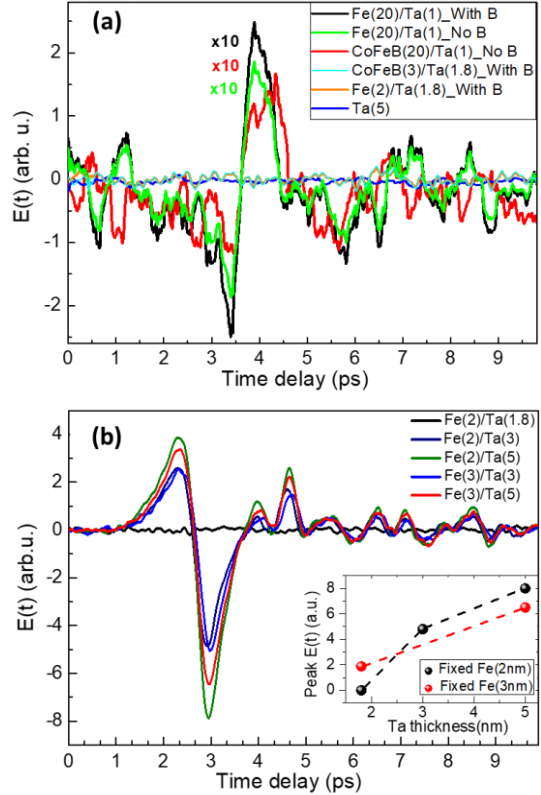

**Figure 2.** (a) THz signal generated from individual FM and NM layers. Very thin Ta layer (thickness ~1.5 nm) can be considered as the capping layer for FM layer. (b) THz signal from Fe/Ta bilayers to obtain optimized thickness of the Fe and Ta layers. Inset: THz amplitude with the increasing thickness of the Ta layer at fixed thickness of the Fe layer (2 nm and 3 nm).

The ultrafast demagnetization is also a process, which, under femtosecond excitation of FM layers, generates THz radiation and has been reported in several studies.[10, 37, 38] From our results presented in Figure 2(a), we intend to disseminate a conclusion that the FM and NM layers individually contribute very weakly to the THz generation process. As discussed later, THz signal produced by ISHE is about two orders of magnitude stronger. From Figure 2(a), we can see that no THz signal is produced at all from Ta(5), Fe(2)/Ta(1.8) and CoFeB(3)/Ta(1.8) samples. Clearly, THz generation is not possible either through ultrafast demagnetization in thin FM layers or through ISHE in FM/NM bilayers when the NM layer thickness is not appropriate. A weak THz generation from thick FM layers through ultrafast demagnetization is evident from our results presented in Figure 2(a) for the Fe(20)/Ta(1) and CoFeB(20)/Ta(1) samples, where the signals have been magnified by a factor 10 for visualization. These results are consistent with the literature.[38] Here, in all the cases, a thin Ta (NM) layer has been utilized as a capping layer (thickness ~1nm), where it is expected that such a thin layer does not play any role to the THz generation because of small spin-Hall coefficient (see Figure 2(b)). We also note from Figure 2(a) that there is no role of the external magnetic field, i.e., the magnitude of the weak THz signal does not change with the external magnetic field.[38, 39] All the results in Figure 1 and 2 have been obtained for a fixed excitation fluence of ~1.2 mJ/cm$^2$.

THz generation by ISHE in thin film Fe/Ta heterostructures can be observed only if the Ta layer thickness is larger than 2 nm. The corresponding results have been presented in Figure 2(b), where, data is shown for Fe layer thicknesses 2 nm and 3 nm, while Ta layer thickness varying from 1.8 nm to 5 nm. These results again confirm that the optimized thickness of the Fe layer for maximum THz signal is ~2 nm. As shown in the inset of Figure 2(b), the THz signal continues to grow with the increasing Ta layer thickness. While comparing the results from the Fe/Pt (Figure 1) and Fe/Ta (Figure 2(b)), we find that the optimum thickness of the Ta layer is much larger than the Pt layer. It is noteworthy from the results discussed above that the demarcation line for the thickness of the α−phase Ta layer below which it can be used as a capping layer is ~2nm. However, it may vary depending on the different phases of the Ta.[36]

The presence of an interface and intermixing at the interface can contribute to the generation of THz signal through various mechanisms, which can be probed through excitation pulse helicity dependent measurements. For example, the THz generation process based on ISHE does not depend on helicity, whereas, the spin dependent photogalvanic effect (SDPE) is dependent on the helicity of the excitation pulse.[1] Our experimental results presented in Figure 3 suggest that only ISHE is responsible for the THz generation from FM/NM heterostructures. The results in Figure 3 have been obtained on optimized Fe(2)/Pt(3) bilayer for linearly polarized (LP), elliptically polarized (EP), left circularly polarized (LCP) and right circularly polarized (RCP) light, where, in each case, the excitation fluence was kept fixes at ~1.2 mJ/cm$^2$. We may also point out that THz generation from antiferromagnetic/nonmagnetic film heterostructures arises due to a different mechanism, where, helicity dependent outcome is obtained.[40] Therefore, the contribution from the SDPE is negligible.

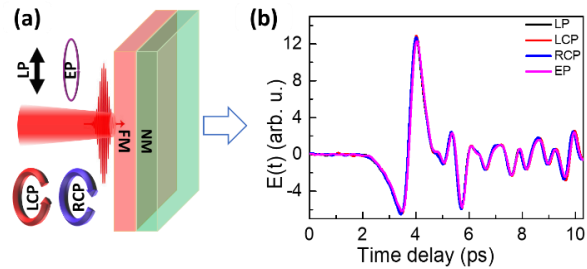

**Figure 3.** Incident laser helicity-dependent THz emission from Fe(2)/Pt(3) bilayer heterostructure. (a) Schematic of the experiment, and (b) corresponding THz time-domain signals. LP: Linear polarization, LCP: Left-handed circular polarization, RCP: Right-handed circular polarization, EP: Elliptical polarization.

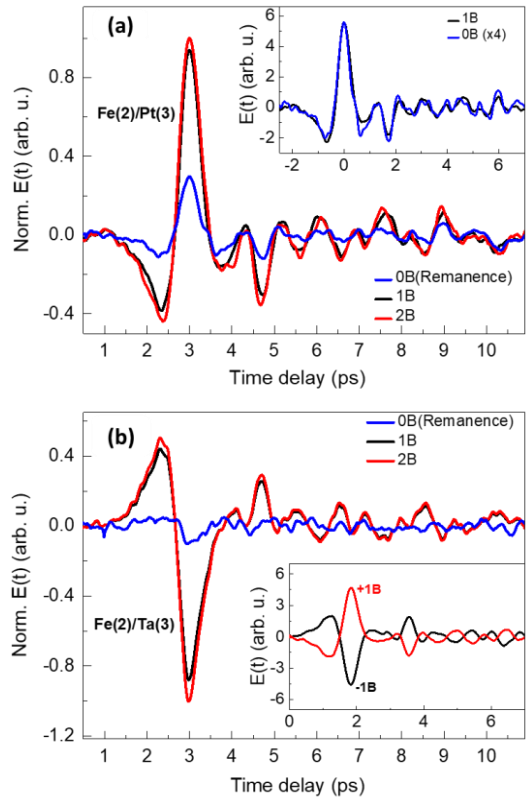

**Figure 4.** Time-domain THz traces normalized by the maximum amplitude from (a) Fe(2)/Pt(3), and (b) Fe(2)/Ta(3) spintronic emitters in the presence of zero (remanence only), 1B, and 2B magnetic field. B denotes the applied magnetic field value of ~200 Oe, which is in the saturation magnetization. Inset: (a) Comparison between the THz time traces in case of applied and no applied magnetic field, and (b) THz signals for +1B and −1B.

As per the Eq. (1), the magnetization of the FM layer can be used as a control to manipulate the amplitude and polarization of the emitted THz radiation.[14, 41] In Figure 4(a) and 4(b) we have shown the THz signals from Fe(2)/Pt(3) and Fe(2)/Ta(3) samples, respectively, with and without the external field, $B_{ext}$ and for the pump fluence fixed at ~1.2 mJ/cm$^2$. For the Fe layer, the saturation magnetic field (coercive field) is ~50 Oe as confirmed from the in-plane M-H measurements (see Supplementary Section S1). Therefore, for the magnetic field effect, we have shown the results for three values, i.e., 0B (no field), 1B ~200 Oe and 2B ~400 Oe. In the first case (0B), only the effect of any remanence field is observed in terms of a small magnitude THz signal,[7] due to the pinned magnetic domains avoiding reorientation after removal of the external field. For the other two field values (1B and 2B), the THz signal is already saturated. Other than the change in the THz magnitude, the temporal profiles of the pulses are exactly the same irrespective of the field value (see inset of Figure 4(a)). As shown in the inset of Figure 4(b), the polarity of the generated THz signal can be reversed by flipping the direction of the external field. This observation is consistent with Equation (1) and hence, clearly demonstrates the ISHE origin for the generation of THz radiation from FM/NM structures.[6]

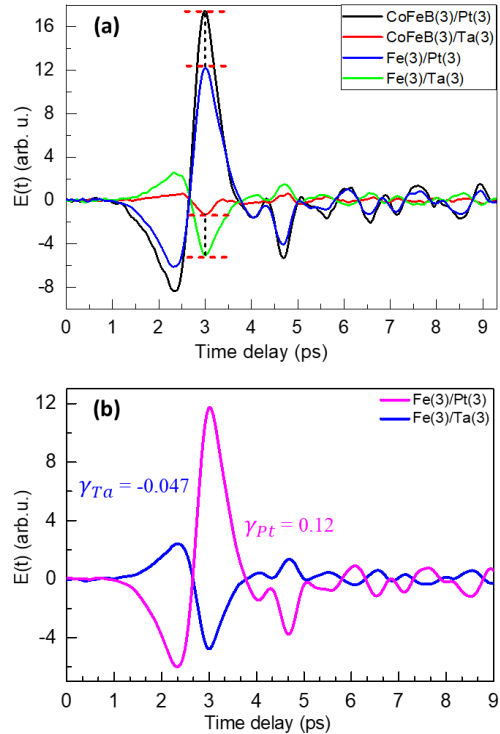

**Figure 5.** (a) Comparison between THz signals generated from Pt and Ta capped Fe and CoFeB heterostructures. (b) Experimental estimation of spin-Hall angle, γ in Fe(3)/Ta(3) bilayer heterostructure.

Now in Figure 5, we compare the THz generation efficiency of the nearly optimized Fe/Pt, Fe/Ta, CoFeB/Pt and CoFeB/Ta bilayer heterostructures having same thicknesses of the FM and NM layers in each of the combinations and under the same experimental conditions. The CoFeB(3)/Pt(3) heterostructure provides ~40% stronger THz emission than that from Fe(3)/Pt(3). The enhanced THz signal from the prior can be attributed to the larger spin injection into the NM layer and is consistent with the literature.[6] Interestingly, the case is entirely reversed for the bilayers containing Ta as the NM layer. Also, the THz signal from Fe(3)/Ta(3) is about 300% stronger than that from CoFeB(3)/Ta(3) as shown in Figure 5(a). The spin injection efficiency in the CoFeB(3)/Ta(3) seems to have got severely affected by the seemingly larger crystalline mismatch and roughness at the FM/NM interface, both of which essentially affect the spin mixing conductance ($g_{\uparrow\downarrow}$) and spin Hall angle at the interface.

The spin mixing conductance, ($g_{\uparrow\downarrow}$) is the measure of the spin current injected from the FM layer into the NM layer. Assuming a simple precession motion, the spin current density is related to the spin mixing conductance through a relation given by[36, 42]: $J_s = (1/2\pi)e\omega\text{Re}(g_{\uparrow\downarrow})\sin(\theta_c)$, where, $e, \omega, \text{Re}(g_{\uparrow\downarrow})$, and $\theta_c$ are electronic charge, angular frequency of spin dynamics, real part of spin mixing conductance, and the magnetization precession angle, respectively. Therefore, a higher value of $\text{Re}(g_{\uparrow\downarrow})$ ensures enhanced THz emission through Eq. (1). The reported values of the spin mixing conductance from literature are: ~3.5-6×10$^{19}$ m$^{-2}$ for CoFeB/Pt,[43-45] ~3-5×10$^{19}$ m$^{-2}$ for Fe/Pt,[46, 47] and ~0.7-1.4×10$^{19}$ m$^{-2}$ for CoFeB/Ta.[44, 45] For the CoFeB/Ta case, Ta in its β-phase was used[44, 45]. Similar value for Fe/Ta could not be found in the literature. Moreover, the FM/NM spintronic heterostructures containing α-phase Ta have not been studied much in the literature. The α-phase of Ta is more metallic than the β-phase and hence can be better suited for the THz applications though it has comparatively a slightly smaller spin Hall coefficient[36]. The experimentally observed small relative change in the THz signal from CoFeB/Pt as compared to Fe/Pt (Figure 5(a)) can be considered consistent with the corresponding values of the $g_{\uparrow\downarrow}$. However, the much larger THz signal from Fe/Ta than the CoFeB/Ta demands that the corresponding $g_{\uparrow\downarrow}$ value in Fe/Ta should be higher than the CoFeB/Ta. Our THz experiments suggest a higher value of $g_{\uparrow\downarrow}$ for the Fe/Ta as compared to CoFeB/Ta. The conclusion of this part of the results is that Pt should be favored with CoFeB, while α-phase Ta should be favored with Fe in their FM/NM bilayer structures for best THz generation performance.

Besides the electronic mechanisms,[13] a comparison in the magnitude of the THz signals from two FM/NM combinations can be used to quantify the spin Hall angle in a given NM layer[15]. For example, the THz generation efficiencies of CoFeB/Pt and CoFeB/W were compared to determine the spin Hall angle in W to be γ ~ -0.056, which was very close to its exact value.[48] We have used the known value of γ = 0.12 in Pt[35, 49, 50] to estimate the same in α-phase Ta by comparing the THz signal strengths from Fe/Pt and Fe/Ta in Figure 5(b) having the same FM layer. It is obtained to be γ ~ -0.047 from the ratio between the RMS values of the corresponding THz signal strengths. Of course, for the exact determination of the γ value, one needs the information about the corresponding $g_{\uparrow\downarrow}$ value and other material parameters, preciously[15]. However, we may note that the estimated value of γ for the α-phase Ta in the above, comes out to be quite close to a value reported in the literature for NiFe/Ta, where Ta was used in the α-phase.[36]

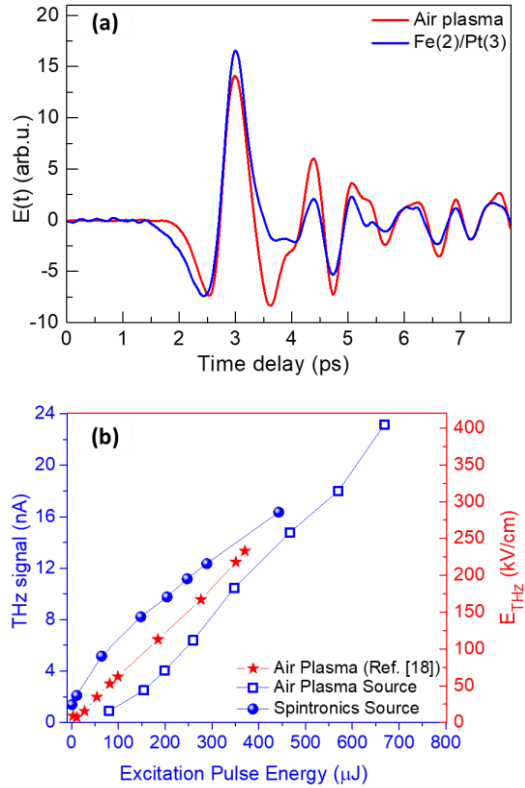

**Figure 6.** (a) Time-domain THz traces from dual color air-plasma source and optimized Fe/Pt spintronic emitter under same excitation power and other optical conditions. (b) Amplitude of the THz signal from air-plasma source (blue squares) and spintronic source (blue spheres) for various excitation pulse energies. The graph for THz electric field (red stars) for an air-plasma source is drawn from Ref. [18], for a comparison.

For a quantitative assessment of the THz generation efficiency of the spintronic emitters, here, we compare the THz signal strength with a more standard high power THz source, i.e., the dual colour air plasma source. To do so, excitation fluence-dependent measurements were carried out on both the sources in configuration I of the experimental set up shown in Figure S2.1 and S2.2 of the supplementary information. In Fig. 6(a), the THz signals from the Fe(2)/Pt(3) and the plasma sources are presented that were taken at same excitation pulse energy of ~350 μJ. It can be seen that the THz emission from both the sources is quite comparable. In fact, the THz generation efficiency of the spintronic source is slightly better than the air-plasma source at low excitation powers as shown in Figure 6(b).

In the absence of a THz power/energy meter, we could estimate the strength of the generated THz signal from our measurements in the units of THz electric field (kV/cm). To do so, we have compared in Figure 6(b), the excitation energy dependent THz signal strengths in nA (lock-in units) from the plasma source (our experiments) with the reported values in the literature for the same source[18] in units of kV/cm. We note that these experiments have been performed under very similar experimental conditions with the excitation pulse duration, the SHG crystal, focussing, etc. Nearly one-to-one correspondence in the two units can be seen in Figure 6(b) for the air-plasma source. Therefore, a THz signal of strength 10 nA in our case can be considered equal to actual THz field strength of ~170 kV/cm. Furthermore, the peak amplitude of the THz electric field from the air plasma source linearly increases with the excitation pulse energy in Fig. 6(b). The linear behaviour is a good indication that electro-optic effect in the ZnTe crystal is still in the linear regime for the detection of such intense THz pulses.[51] This is because of the excitation energy values being much above the air plasma ionization threshold value[18, 52] of ~$10^{14}$ W/cm$^2$. A slight difference in the two graphs in Figure 6(b) can be attributed to the differences in the experimental conditions due to room temperature, humidity, and the experimental setup.

As mentioned before, for reasonable excitation power, the spintronic source produces slightly higher THz signal than the air plasma source (Figure 6(a)) and it grows nearly linearly with the excitation power. However, at very high values of the excitation power, the spintronic sources will not work because of the optical damage to the material. Here, we come to the second main point of our paper, i.e., to determine the optical damage threshold of various spintronic THz emitters used in our study. The excitation power was varied in a large range by using a neutral density filter in the excitation path, while keeping the excitation beam size fixed at ~1 mm on the sample. These measurements were done in different configuration of the experimental setup as shown in Figure S3 of the supplementary information. In this case, the spintronic source was placed at the focal point between the two inner parabolic mirrors and a converging optical beam through the hole of one of the parabolic mirror irradiated the sample. For the Fe(2)/Pt(3) heterostructure on the quartz substrate, the THz peak amplitude with respect to the excitation fluence (mJ/cm$^2$) has been presented in Figure 7(a). The THz power continues to grow upto excitation fluence of ~5.5 mJ/cm$^2$ beyond which a sudden decrease in the THz signal can be noticed. This is due to the optical damage of the spintronic material. Optical microscopy images of the sample, before and after the optical damage are shown in the inset of Figure 7(a) (please see the Supplementary Information S4 for more details).

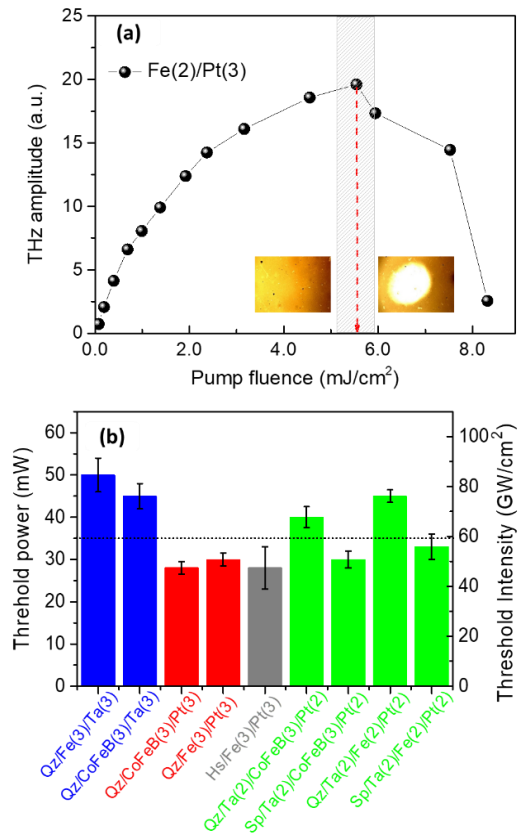

**Figure 7.** (a) Peak THz amplitude with the increasing pump fluence of a FM/NM spintronic heterostructure. The vertical mesh shows the region of damage occurrence and the dashed line indicates the damage threshold value. The sample's microscopy image before and after the threshold power. (b) Threshold power and corresponding peak intensity of various spintronic emitters. The number following the material layer indicates the film thickness in nm. Qz: Quartz; Sp: Sapphire; Hs: Highly resistive silicon.

A weak saturation behaviour of the THz signal with the excitation fluence, much before the optical damage starts, can be seen in Figure 7(a). The saturation can arise due to mainly two reasons, the spin accumulation, and the local heating.[53, 54] If the spin accumulation induced saturation[5, 7] was stronger than the later, we should have observed saturation in the signal quite early in the data shown in Figure 7(a). Therefore, the local heating induced saturation in the THz signal is more dominant in our case. We may note that the irradiation time at each fluence was about 5 minutes.

The optical damage threshold (ODT) of the spintronic emitters depends on the FM/NM layer combinations in bilayers and trilayers, their material types, and the underlying substrates. To investigate this aspect in detail, we have carried out experiments on multiple samples and plotted the corresponding ODT values in mW (average power) and $GW/cm^2$ (peak intensity) in Figure 7(b). More details of the experiments are given in Section S4 of the Supplementary information. In Figure 7(b), the samples have been named as per the substrate used and the bi- or the tri-layer heterostructure having the individual layer thicknesses given inside small parentheses. We note that for each of the bilayer and the trilayer, we have used the optimized thickness of the FM and the NM layers. Also, the overall material thickness in both the bi-and tri-layer combinations have been kept the same. The substrates used are: 1 mm thick quartz (Qz) plate, 1 mm thick sapphire (Sp) plate, 0.38 mm thick highly resistive silicon wafer (Hs). The mean value of the ODT for the spintronic THz emitters is about 35 mW and corresponding peak intensity is ~60 $GW/cm^2$. This value is only a few times smaller than the value (~100 $GW/cm^2$) for the popular lithium niobate crystal, which has been used in the literature[22, 55] as a high-power THz source, routinely.

From Figure 7(b), we notice that the bilayers on quartz substrate with Ta as the NM layer show the highest ODT value of ~50 mW (~85 $GW/cm^2$). Therefore, quartz is suggested to be a good substrate for better heat sink management to achieve higher ODT in the spintronic THz emitters. It appears from Figure 7(b) that the ODT value is nearly independent of the type of the FM layer but affected by the type of the NM layer in the bilayer heterostructures on the same type of substrate. The reason for this fact is related to the overall absorptance and reflectance/transmittance of the structures as shown in Figure S4 of the Supplementary information. The spintronic heterostructures on sapphire and HR silicon show the least ODT value because of the higher optical absorption in those substrates. As shown in Fig. S4 of the supplementary information, the ODT values also depend on the incidence angle of optical excitation, and it is higher for larger angles.

3. Conclusion

To summarize, we have studied the THz generation efficiency of spintronic THz emitters limited by the optical damage threshold at 800 nm wavelength. Such a study is very important from the point of view that in recent literature on the spintronic THz emitters, they have been suggested to be the highly efficient, powerful, and broadband sources. We have explored the contributions from the ultrafast demagnetization process in the ferromagnetic layers and observed that the inverse spin Hall effect is the main origin for high power THz generation from the bi- and tri-layer FM/NM type spintronic THz emitters. Detailed experiments were also performed to obtain the optimized thicknesses of the individual FM and NM layers in their bi-layer combinations. It is shown that the THz emission is insensitive to the polarization and helicity of the excitation laser light. The THz power from these sources grows nearly linearly with the excitation power before the optical damage threshold of the material heterostructure. It has been found that Fe/Ta on the quartz substrate has the highest optical damage threshold of ~85 $GW/cm^2$. From the many spintronic systems studied, it was realized that the mean value of the optical damage threshold for such THz emitters is about 50 $GW/cm^2$. These results are highly encouraging and suggest that suitable materials and their heterostructures, which provide simultaneously, high optical damage threshold and high THz generation efficiency, can be discovered.

4. Experimental Section

*Sample Preparation and characterization*: High quality bilayer and tri-layer heterostructures of thin films containing ferromagnetic (FM) and nonmagnetic (NM) metallic materials were developed using radio frequency (RF) magnetron sputtering (AJA international). The lateral size of the films was larger than 10 mm x 10 mm and they were deposited on different substrates, 1 mm quartz, 1 mm sapphire and 0.38 mm highly resistive silicon. The base pressure for the thin film growth was kept at $5\times10^{-8}$ Torr, while the working pressure was kept in the range of 2-5 mTorr according to the optimization condition for each of the deposited material film. Prior to deposition, the substrates were cleaned by double sonication in acetone + isopropyl alcohol (IPA) to remove the contamination from their surfaces. To achieve high purity samples, we used high purity targets and pre-sputtered them to get rid of the impurities on the target surfaces before the desired film deposition. To ensure good uniformity, the substrate holder was kept rotating with a constant speed during the deposition. Different thickness combinations in the bilayer FM/NM and tri-layer $NM_1$/FM/$NM_2$ heterostructures were created using Fe and $Co_{20}Fe_{60}B_{20}$ (CoFeB) as ferromagnetic materials and platinum (Pt) and α-phase tantalum (Ta) as nonmagnetic materials at a growth rate of 0.09 Å/sec, 0.25 Å/sec, 0.44 Å/sec, and 0.76 Å/sec, respectively. The RF power was also kept according to the optimized phase of the deposited material, i.e., 150 Watts for Ta and 100Watts for Fe, CoFeB and Pt. The samples or the substrates were not treated with any pre- and/or post-annealing processes.

The structural and magnetic characterizations of the deposited thin film samples were obtained from X-ray diffraction (XRD), and magnetic hysteresis (M-H) measurements, respectively (see Section S1 of the Supplementary information). The XRD results confirm the polycrystalline phase of Pt and α-phase of Ta. The surface and interfacial roughness from the XRR fitting were found to be <0.5 nm for all the samples. The in-plane magnetization of the Fe and CoFeB films in Fe/Pt and CoFeB/Pt bilayer heterostructures is confirmed by performing in-plane and out-of-plane M-H measurements. A clear square hysteresis loop with the saturation magnetic field value of <50 Oersted (Coercive field) is obtained for the magnetic layers. Therefore, the samples were kept magnetized along certain direction by using external magnetic field value of >50 Oe during the THz generation from the spintronic heterostructures.

*THz time-domain spectroscopy*: A time-domain THz spectrometer (TDTS) was developed around a Ti:sapphire regenerative amplifier, operating at 1 kHz pulse repetition rate and providing <50 fs laser pulses centered at 800 nm. Dispersion uncompensated pulses were used and the pulse duration at the sample point was ~100 fs. The layout and other details of the setup in different configurations of the measurements are provided in Figure S2.1, S2.2 and S3 of the supplementary information. The detection of the THz pulses was made on a [110] oriented 500-micron thick ZnTe crystal by electro-optic sampling. Collimated excitation beam (size ~6 mm) at a fixed pulse energy (fluence) of ~350 μJ (1.2 mJ/cm$^2$) was used to irradiate the spintronic emitters in the configuration 1 of the setup for determining optical thickness of the FM and NM layers and their combinations in the heterostructures for best THz power generation efficiency. For quantification of the THz power from the spintronic emitters, we developed a dual color air-plasma based THz source (see Figure S2.2 of the supplementary information). The excitation pulse energy dependent THz output of our air-plasma source was compared with that from the literature for a similar setup and optical conditions with the excitation pulse and detection. For determining the optical damage threshold of the spintronic emitters, we used the same experimental setup in configuration 3 as shown in Figure S3. Here, a converging excitation beam passing through the aperture of one of the parabolic mirrors is used for optical pumping of the source and excitation power was controlled by using a neutral density filter. All the experiments, either with a fixed excitation fluence or with varying fluence, have been performed under the same experimental conditions with the room temperature, humidity (~50%), excitation pulse duration, optical alignment, detection, etc. Therefore, comparisons among different experiments for relative performance parameter was possible.

**Supporting Information**
Supporting Information is available.

Acknowledgements

SK acknowledges Science and Engineering Research Board (SERB), Department of Science and Technology, Government of India, for financial support through project no. CRG/2020/000892, Joint Advanced Technology Center, IIT Delhi and Director, LASTEC Delhi are also acknowledged for support through EMDTERA#5 project. Mr. Akash Kumar and Dr. P.K. Muduli of the Physics Department are thanked for help during the material synthesis using UHV sputtering. One of the authors (Sandeep Kumar) acknowledges the University Grants Commission, Government of India for Senior Research Fellowship.

# Supporting Information

# Optical Damage Threshold and THz Generation Efficiency of (Fe,CoFeB)/(Ta,Pt) Spintronic Emitters

**Sandeep Kumar, Anand Nivedan, Arvind Singh, Yogesh Kumar, Purnima Malhotra, Marc Tondusson, Eric Freysz, and Sunil Kumar**[*]

*\*kumarsunil@physics.iitd.ac.in*

**Content:**
S1.   Structural and magnetic characterizations of the samples
S2.   THz time-domain spectroscopy setup
S3.   Optical excitation power-dependent measurements on the spintronic emitters
S4.   Optical damage threshold of spintronic THz emitters

**S1. Structural and magnetic characterization of the samples**

X-ray diffraction (XRD) and X-ray reflectivity (XRR) measurements were carried out using a PANalytical X'Pert diffractometer with a Cu-K$_\alpha$ source to determine the crystalline phase and purity of Pt and Ta thin films (Figure S1 (a) and (b)). In case of Pt, various XRD peaks have been marked with the corresponding crystallographic planes and confirm the polycrystalline phase of the film that is consistent with the literature.[1] In the XRD pattern of Ta film, a broad diffusive peak centred at angle $2\theta = 38.0^0$ is due to the Bragg reflection from (110) planes and evidently shows the growth of Ta primarily in its α-phase. The same could not be done for the ferromagnetic thin films due to the limitation with the X-ray source and rapid oxidation problem with these uncapped films[2]. The crystalline α-phase of Ta is deliberately obtained by controlling the growth parameters during the deposition. The thickness and surface/interface roughness of the fabricated samples were found to be consistent with those expected from the optimized growth rate parameters in the deposition, which are confirmed from the XRR data fitted using the recursive theory by Parratt[3]. The surface and interfacial roughness is found to be less than ~0.6 nm for our samples.

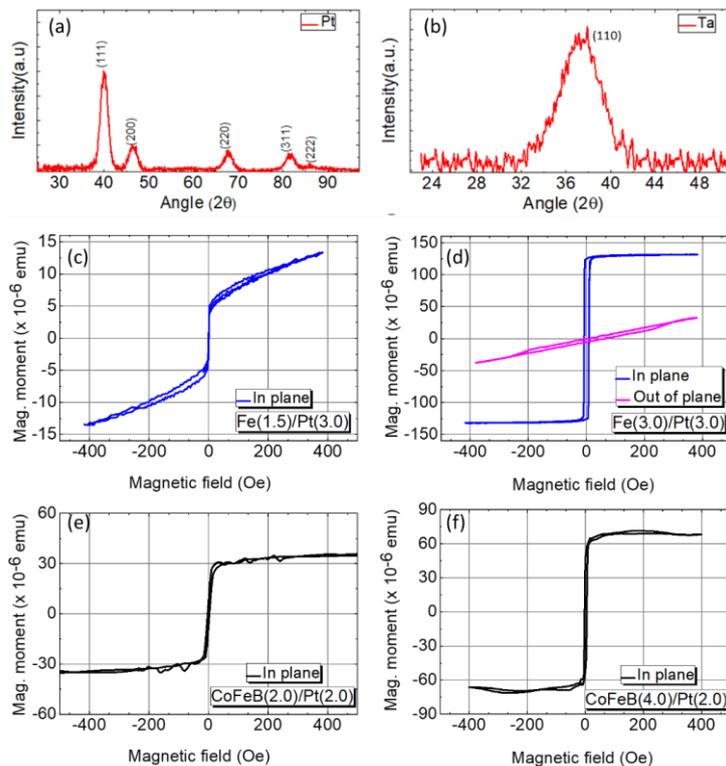

**Figure S1** XRD plots of (a) Pt film, and (b) α-phase Ta film. The crystallographic planes corresponding to various XRD peaks have been marked. Hysteresis loops from in-plane and out-of-plane magnetic measurements using vibrating sample magnetometer on (c) Fe(1.5)/Pt(3.0), (d) Fe(3.0)/Pt(3.0), (e) CoFeB(2.0)/Pt(2.0), and (f) CoFeB(4.0)/Pt(2.0) bilayer FM/NM spintronic heterostructures.

Since the magnetization behaviour of the ferromagnetic layer in the FM/NM structures, plays a crucial role in the generation of the THz radiation[4], hence, the in-plane/out-of-plane magnetic hysteresis measurements were performed using vibrating sample magnetometer (VSM) of a physical property measurement system (PPMS from Quantum Design) as shown in Figure S1(c) to S1(f). The magnetic hysteresis loops are obtained from the in-plane and out-of-plane M-H measurements on Fe/Pt and in-plane for CoFeB/Pt bilayer heterostructures having fixed Pt layer thickness, but different thicknesses of the Fe and CoFeB layers. It can be seen from the Figure S1(c) and (d) that for Fe layer thickness of 1.5 nm, it shows a non-saturating type of behaviour, whereas, for higher thickness, the sample shows a clear square hysteresis with the saturation magnetic field value of <30 Oersted (Coercive field). The results for CoFeB/Pt show a saturating behaviour for both the thicknesses of the CoFeB layer, while the corresponding coercive field value is ~20 Oe. The change in the nature of the M-H curve from nonsaturating to saturating, while increasing the Fe layer thickness from 1.5 nm to 3 nm, is because of the decrease in the relative contribution from the out-of-plane component of the magnetization. Clearly, for the 1.5 nm thickness, the magnetic easy-axis has a finite component in the out-of-plane, which vanishes above a critical thicknesss.[5] Therefore, for Fe, the critical thickness is in between 1.5 nm and 3 nm, an observation that is consistent with the previous reports.[6] Our results clearly suggest the synthesis of high quality thin films by RF magnetron sputtering.

In our THz generation measurements, the applied external in-plane magnetic field has been kept well above the saturation magnetization, i.e., $B_{ext}$ > coercive field value. It may be pointed out that the out of plane magnetic field is not important for our THz experiments as the charge current responsible for generation of THz radiation is insensitive to the out of plane magnetization direction. Therefore, for all the comparative studies performed in our paper are for FM layer thicknesses such that the magnetic easy-axis lies in the plane, and for the external magnetic field lying in the surface/plane of the layer.

**S2. THz time-domain spectroscopy setup**

For testing the THz generation performance of our spintronic emitters, we developed a time-domain THz spectrometer (TDTS) that was pumped by ~35 femtosecond (fs) laser pulses centred at 800 nm from a Ti:sapphire regenerative amplifier operating at a 1 kHz pulse repetition rate. The optical layout of the TDTS is shown in Figure S2.1, S2.2 and S3 used in three different configurations. In configuration 1 (Figure S2.1), the spintronic emitters are used. The open setup is not purged with dry air/nitrogen, rather, all the experiments for our paper have been performed under normal humidity (~50%) at room temperature of the lab. The laser beam is divided into two-part in a 90:10 beam splitter: the stronger part (pulse energy ~350 μJ and beam diameter ~6mm) is used to pump the THz emitter (spintronic heterostructure), while the weaker part (pulse energy ~0.4 μJ) was routed through a linear translational stage (for generating computer-controlled time-delays relative to the pump/excitation pulse) and used as the gating beam for the detection of the THz pulse on a nonlinear crystal. The diameter of the gating beam was kept at ~2 mm using an iris.

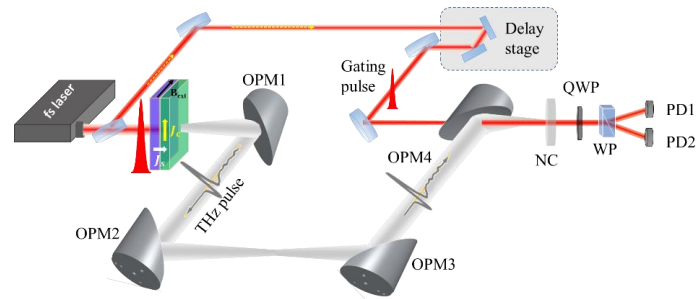

**Figure S2.1** Configuration 1: Time-domain THz spectroscopy setup using spintronic emitters. Pulsed THz generation from the excitation of the spintronics heterostructures with femtosecond NIR pulse and probing of the THz pulses by electro-optic sampling in a nonlinear optical crystal. OPMs: off-axis parabolic mirrors, QWP: quarter-wave plate, NC: nonlinear optical crystal, WP: Wollaston prism, PD: photodiode, LIA: lock-in amplifier, $B_{ext}$: applied external magnetic field, $J_s$: spin current, $J_c$: charge current.

The collimated pump beam excites the spintronic emitter as shown in Figure S2.1. The excitation pulse energy or power is controlled by using a neutral density filter. THz radiation from the spintronic source is collected by using a 15 cm focal length 90⁰ off-axis gold-coated parabolic mirror (OPM1). There are 3 more same type OPMs used in our setup for giving a flexibility to the setup for different experiments[7, 8]. After routing the THz beam through the four OPMs, it is focused onto a (110)-surface oriented 0.5 mm thick ZnTe crystal, where, it collinearly matches with the gating pulse in space and time. A high resistive silicon (HR-Si) wafer is placed just before OPM1 to avoid the residual 800 nm pump laser beam completely from reaching the other side of the setup. Commercial neodymium magnets were used to produce uniform and static magnetic field ($B_{ext}$) of strength ~120 mT. Detection of the THz pulses takes place on the ZnTe crystal via electro-optic sampling of the gating pulse[7, 8] and achieved using a quarter-wave plate, Wollaston prism, a balanced photodiode and lock-in amplifier. The pump beam was modulated at a frequency of 267 Hz by an optical chopper.

For a comparison between the THz generation efficiency of our spintronic emitters and the more standard technique based on a dual-colour air plasma source, we reconfigured our TDTS setup for the later. The layout of the setup in this configuration (configuration 2) is shown in Figure S2.2. Of course, in this case, only the THz generation parts is modified, while all other optical conditions with the pulse duration, optical alignment, and detection by electro-optic sampling are left unchanged. Therefore, while estimating the THz pulse energy/power from our air-plasma source by comparing it with the reported data in the literature for similar amplified laser pulse duration and pulse energies, any differences can be related to the optical conditioning in the two experiments. For the generation of intense THz pulses from our dual colour air-plasma source, we used a biconvex lens of focal length 15 cm to focus the fundamental and second harmonic beam in the air. Thus, generated white plasma creates forward propagating THz pulses if the fundamental and the second harmonic beams are mixed together properly by adjusting the orientation and crystal angle. We used a β-barium borate (BBO) type-I SHG crystal of thickness 100 µm for generating the second harmonic beam as shown in Figure S2.2. Thus, emitted THz radiation is collected by the OPM1 and then detected in the ZnTe crystal by electro-optic sampling as describe before.

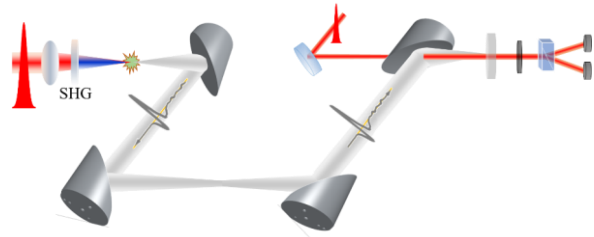

**Figure S2.2** Configuration 2: Time-domain THz spectroscopy setup using dual colour air-plasma source. SHG: second harmonic generation crystal.

## S3. Optical excitation power-dependent measurements on spintronic THz emitters

For doing excitation power-dependent measurements on the spintronic heterostructures for evaluating their THz generation efficiency, we modified the setup to configuration 3 as shown in Figure S3. Here, the spintronic source is placed between the two inner parabolic mirrors and irradiated with converging excitation beam allowed through the aperture in the second OPM. The diameter of the optical beam on the sample was ~1 mm. Like before, no change in the optical conditioning with all other things including the electro-optic detection, we made. Again, a HR-Si wafer was placed just after the emitter to avoid the residual pump from reaching the second part of the setup.

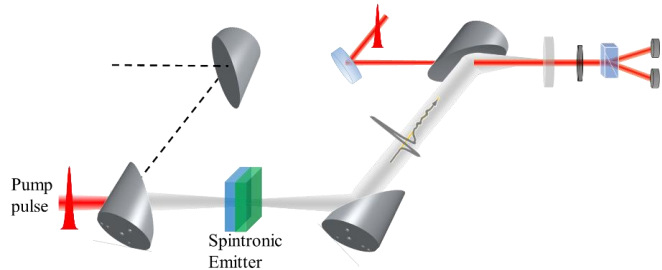

**Figure S3** Configuration 3: Experimental setup for THz time-domain spectroscopy using spintronic emitters and to determine their optical damage threshold.

## S4. Optical damage threshold of spintronic THz emitters

Figure S4.1 shows the optical microscopy images of a spintronic source before and after the optical damage. The images are taken in the transmission mode using a 10X objective lens. The sample for representation is Fe(2)/Pt(3) on quartz substrate and these experiments were conducted using the setup in its configuration 3 as described above. The size of the damage due to excessive heating at high level optical excitation is close to the predetermined size of the excitation beam. The excitation power was continuously increased using a neutral density filter and at few values, the magnitude of the THz signal also was recorded, before a sudden decrease in the signal was observed due to the optical damage. Experiments on various samples on different substrates were conducted for evaluating the performance of spintronic THz emitters with highest THz generation performance and highest optical damage threshold.

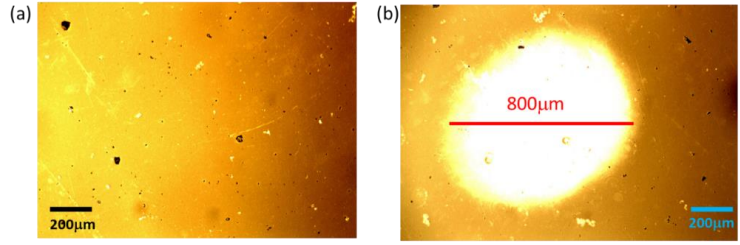

**Figure S4.1.** Optical microscopy images of the representative spintronic emitter taken in the transmission mode using a 10X objective lens. (a) Before, and (b) after the optical damage.

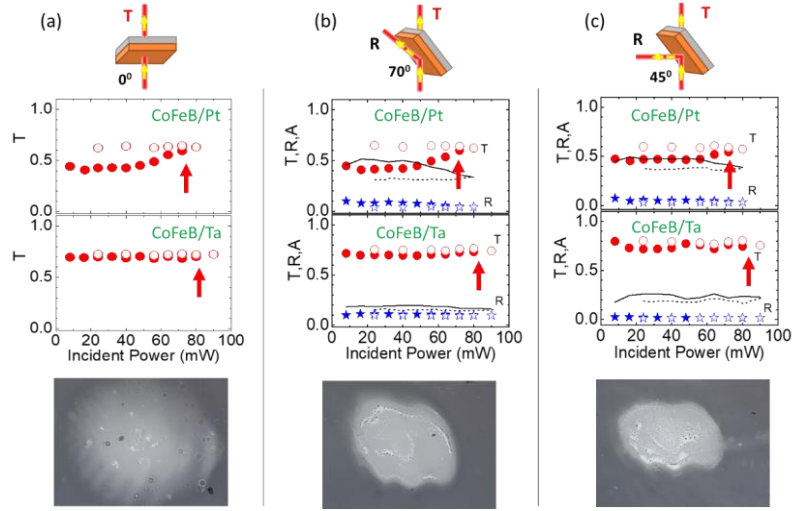

**Figure S4.2** Optical transmittance (T) and reflectance (R) measurements with respect to the incident laser power on the spintronic emitters at three different incident angles. The representative samples are CoFeB(3)/Pt(3) and CoFeB(3)/Ta(3) on quartz substrates. Solid symbols are for increasing power while the open symbols are for lowering power values and the upward arrows in the middle row mark the values of the damage threshold. The images in the bottom row are optical images of the samples after the damage.

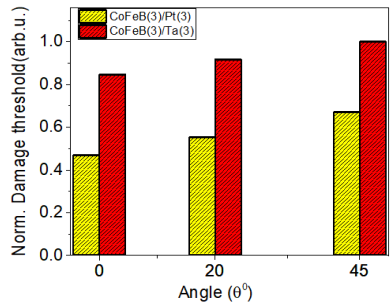

**Figure S4.3** Angle of incidence dependent variation in the damage threshold for CoFeB(3nm)/Pt(3nm) and CoFeB(3nm)/Ta(3nm) bilayer heterostructures.

The optical damage threshold measurements were also done for incident angles different from the normal incidence. The summary of the results on two representative samples, CoFeB(3)/Pt(3) and CoFeB(3)/Ta(3) is given in Figure S4.2 and S4.3 From the amount of reflected and transmitted intensities with respect to the incident intensity, we calculated the reflectance (R) and transmittance (T) of the samples after correcting them by the underlying quartz substrate. The transmittance of CoFeB(3)/Pt(3) is smaller than that of the CoFeB(3)/Ta(3). The absorptance (A) can be calculated using the relation, $A = 1 - R - T$, and they have been shown by continuous and dotted curves in Figure S4.2(b) and (c). The R, T, and A measurements have been done initially for an increasing pump power and then continuously decreasing it from the critical value at which the damage has occurred. The optical damage threshold value (power) of the samples is marked by an upward arrow indicator in the R-T graphs of Figure S4.3. The damage threshold of the CoFeB(3)/Ta(3) is higher than that for CoFeB(3)/Pt(3). For the excitation wavelength of 800 nm, the lower value of absorptance of CoFeB(3)/Ta(3) than CoFeB(3)/Pt(3) is consistent with the higher damage threshold for the prior, as shown in Figure S4.3.